\documentclass[12pt]{article}

\usepackage{graphicx,amsmath,amssymb}
\usepackage{ulem}
\usepackage{bm}
\usepackage{hyphenat}

\usepackage[linktocpage=true]{hyperref}
\hypersetup{colorlinks=true, 
linkcolor = [rgb]{0,0.08,0.45}, 
citecolor = [rgb]{0,0.08,0.45}, 
urlcolor = [rgb]{0,0.08,0.45}}

\newcommand{\mbf}[1]{\mathbf{#1}}
\newcommand{\half}{\textstyle \frac{1}{2}}
\newcommand{\Scal}{\mathcal{S}}
\newcommand{\req}[1]{(\ref{#1})}
\newcommand{\lb}{\label}
\newcommand{\nn}{\nonumber}

\normalsize

\begin{document}

\begin{center}
{\bf {\LARGE Emergent phenomena in QCD: \\  \vspace{6pt} The holographic perspective}\footnote{Invited contribution to the 25th Workshop, ``What Comes Beyond the Standard Models", Bled, Slovenia, July 4--10, 2022.}}
\vspace{.2cm}
\end{center} 

\vspace{0.4cm}
\centerline{{Guy~F.~de~T\'eramond},${}^{\rm a,}$\footnote{\href{mailto:guy.deteramond@ucr.ac.cr} {\texttt{guy.deteramond@ucr.ac.cr}}}}

\vspace{.25cm} 

 \begin{center}
 \it ${}^{\rm a}$Laboratorio de F\'isica Te\'orica y Computacional, Universidad de Costa Rica, \\ 11501 San Jos\'e, Costa Rica
 \end{center}
 
 \vspace{.25cm}

\centerline{\small\today}

 \vspace{.50cm}

\begin{abstract} 

A basic understanding of the relevant features of hadron properties from first principles QCD has remained elusive,  and should be understood as emergent phenomena  which depend critically on the number of dimensions of physical spacetime.  These properties include the mechanism of color confinement, the origin of the hadron mass scale, chiral symmetry breaking and the pattern of hadronic bound states.  Some of these complex issues have been recently addressed in an effective computational framework of hadron structure based on a semiclassical approximation to light-front QCD, and its holographic embedding in anti-de Sitter (AdS) space. This framework also embodies an underlying superconformal algebraic structure which gives rise to the emergence of a mass scale within the superconformal group: It determines the effective confinement potential of mesons, baryons and tetraquarks. The normalizable zero-mass ground state is identified with the pion  which has no baryonic supersymmetric partner, a property which is shared for the lowest meson state in each particle family.  This new approach to hadron physics leads to relativistic wave equations, similar in their simplicity to the Schr\"odinger equation in atomic physics, and provides important insights into the confinement mechanism of QCD.

\end{abstract}

\newpage

\tableofcontents

\flushbottom

\section{Introduction} 

The interactions between the fundamental constituents of hadrons, quark and gluons, observed in high energy  scattering experiments is described to high precision by Quantum Chromodynamics (QCD), establishing this theory as the standard model of the strong interactions. At large distances, however,  the nonperturbative nature of the strong interactions becomes dominant and a basic understanding of the essential features of hadron physics from first principles QCD has remained an important unsolved problem in the standard model of  particle physics. Hadronic characteristics are not explicit properties of the QCD Lagrangian but emergent phenomena, notably,  the origin of the hadron mass scale, the mechanism of color confinement, the relation between chiral symmetry breaking and confinement, the massless pion in the chiral limit (the limit of zero quark masses), bound states and the pattern of hadron excitations. Other important aspects of the strong interaction, such as the emergence of Regge theory, Pomeron physics and the Veneziano amplitude, were introduced in dual models before the advent of QCD, and should also be considered  large distance QCD emergent phenomena.  Our present goal is trying to understand how emerging QCD properties would appear in an effective computational framework of hadron structure, as well as their dependence on the dimensionality of physical spacetime.

QCD admits an Euclidean lattice formulation~\cite{Wilson:1974sk} which has been established as a rigorous framework to study hadron structure and spectroscopy nonperturbatively.  However,  dynamical observables in Minkowski spacetime cannot be obtained directly from the Euclidean lattice.  Quantum computation of relativistic field theories using the Hamiltonian formalism in light-front (LF) quantization~\cite{Dirac:1949cp} represents a promising venue, but its development is still at the exploratory phase~\cite{Kreshchuk:2020dla}.  Other nonperturbative methods based on the Schwinger-Dyson and the Bethe-Salpeter equations, and other approximations and models of the strong interactions, are described in Ref.~\cite{Gross:2022hyw}.

Recent theoretical developments based on AdS/CFT -- the correspondence between classical gravity in a higher-dimensional anti-de Sitter (AdS) space and conformal field theories (CFT) in physical space-time~\cite{Maldacena:1997re, Gubser:1998bc, Witten:1998qj}, have provided a semiclassical approximation for strongly-coupled quantum field theories, giving new insights into nonperturbative dynamics~\cite{Aharony:1999ti}. This approach provides useful tools for constructing dual gravity models in higher dimensions which incorporate confinement and basic QCD properties in physical spacetime.  The resulting gauge/gravity duality is broadly known as the AdS/QCD correspondence, or holographic QCD.

Our approach to holographic QCD is based on the holographic embedding of Dirac’s relativistic {\it front form} of dynamics~\cite{Dirac:1949cp}  into AdS space, thus its name Holographic Light-Front QCD (HLFQCD).  This framework leads to relativistic wave equations in physical space-time, similar to the Schr\"odinger or Dirac wave equations in atomic physics~\cite{deTeramond:2008ht, deTeramond:2013it, Brodsky:2014yha}. This approach has its origins in the precise mapping between the hadron form factors in AdS space~\cite{Polchinski:2002jw} and physical spacetime, which can be carried out for an arbitrary number of quark constituents~\cite{Brodsky:2006uqa}: It leads to the identification of the invariant transverse impact variable $\zeta$ for the $n$-parton bound state in physical 3+1 spacetime with the holographic variable $z$, the fifth dimension of AdS.

A remarkable property of  HLFQCD is the embodiment of a superconformal algebraic structure which is responsible for the introduction of a mass scale within the algebra~\cite{deAlfaro:1976vlx, Fubini:1984hf, Akulov:1983hjq} and leads to a well-defined  vacuum, the lowest  normalizable state~\cite{deAlfaro:1976vlx}.  When extended to holographic QCD, this symmetry determines uniquely the form of  the light-front effective interaction, and the corresponding modification of AdS space. The underlying superconformal structure,  gives rise to a massless pion in the chiral limit and to striking connections between the spectrum of mesons, baryons and tetraquarks~\cite{Brodsky:2013ar, deTeramond:2014asa, Dosch:2015nwa}, thus providing important insights into the QCD confinement mechanism. Further extensions of HLFQCD provide nontrivial relations between the dynamics of form factors and quark and gluon distributions~\cite{deTeramond:2018ecg, Liu:2019vsn, deTeramond:2021lxc} with pre-QCD nonperturbative approaches such as Regge theory and the Veneziano model.

In this introductory presentation I will give an overview of relevant aspects of the semiclassical approximation to QCD quantized in the light front  in 1 + 1 and  3~+~1 spacetime dimensions. I will briefly discuss the holographic embedding in AdS$_5$ space of the 3~+~1 semiclassical QCD wave equations for arbitrary spin, with an emphasis on the underlying  superconformal structure for hadron spectroscopy. Other relevant aspects and applications of the light-front holographic approach to hadron physics have been described in the recent review~\cite{Gross:2022hyw}.


\section{Critical role of spacetime dimension on QCD emergent phenomena}

The number of dimensions of physical spacetime is critical in determining whether hadronic properties are complex emergent phenomena, which arise out of the QCD Lagrangian, or can (at least in principle) be computed and expressed in terms of the basic parameters of the QCD Lagrangian~\cite{deTeramond:2022int}. 

Our starting point is the QCD action in $d$ dimensions with an $SU(N)$ Lagrangian written in terms of the fundamental quark and gluon gauge fields, $\psi$ and $A$,
\begin{align}  \lb{L}
S = \int d^dx \left( \bar \psi \left( i \gamma^\mu D_\mu - m\right) \psi 
- \tfrac{1}{4} G^a_{\mu \nu} G^{a \, \mu \nu} \right),
\end{align}
where $D_\mu = \partial_\mu - i g T^a A^a_\mu$ ~ and ~
$G^a_{\mu \nu} = \partial_\mu A^a_\nu - \partial_\nu A^a_\mu  + {f^{abc} A^b_\mu A^c_\nu}$, with $[T^a, T^b] =  i f^{abc}$ and $a, b, c$ are $SU(N)$ color indices. A simple dimensional analysis of the QCD action gives
\begin{align}
[\psi]  & \sim M^{(d-1)/2}, \\
[A]     & \sim M^{(d-2)/2}, \\
[g]     & \sim M^{(4-d)/2}. \lb{g}
\end{align}

It follows from \req{g} that in 1 + 1dimensions, for example, the QCD coupling $g$ has dimensions of mass,  $[g] \sim M$. In this case, the theory can be solved for any number of constituents and colors using discretized light-cone quantization (DLCQ) methods ~\cite{Pauli:1985pv, Hornbostel:1988ne}: All physical quantities can be computed in terms of the basic 1 + 1 Lagrangian parameters, the coupling and the quark masses, thus no emergent phenomena appear.

In contrast, in 3+1 dimensions the coupling $g$ is dimensionless and, in the limit of massless quarks, the QCD Lagrangian is conformally invariant\footnote{The QED abelian coupling $\alpha = e^2/4 \pi$ is also dimensionless in 3 + 1 dimensions, but the physical observables in atomic physics can be directly computed  and, in contrast with the proton, depend critically on the constituent masses in the QED Lagrangian.}.  The need for the renormalization of the theory introduces a scale $\Lambda_{\rm QCD}$, which breaks the conformal invariance and leads to the ``running coupling" $\alpha_s\left(\mu^2\right)= g^2(\mu)/ 4 \pi$ and asymptotic freedom~\cite{Gross:1973id, Politzer:1973fx} for large values of  the momentum transfer  $\mu^2$.  The scale $\Lambda_{\rm QCD}$ is determined in high energy experiments: Its origin and the emergence of the hadron degrees of freedom out of the constituent quark and gluon degrees of freedom of the QCD Lagrangian remains a deep unsolved problem. 


\section{Semiclassical approximation to light-front QCD}

LF quantization uses  the null plane $x^+ = x^0 + x^3 = 0$ tangent to the light cone as the initial surface,  thus without reference to a specific Lorentz frame~\cite{Dirac:1949cp}. Evolution in LF time $x^+$ is given by the Hamiltonian equation
\begin{align} \lb{LFHE}
i \frac{\partial}{\partial x^+} \vert \psi \rangle = P^- \vert \psi \rangle, \quad  \quad
 P^- \vert \psi \rangle = \frac{\mbf{P}_\perp^2 + M^2}{P^+}  \vert \psi \rangle, \quad 
\end{align}
for a hadron with 4-momentum  $P =  (P^+, P^-, \mbf{P}_{\!\perp})$, $P^\pm = P^0 \pm P^3$, where the LF Hamiltonian $P^-$ is a dynamical generator and $P^+$  and $\mbf{P}_{\!\perp}$ are kinematical.  Hadron mass spectra can be computed from the LF invariant Hamiltonian ~$P^2  =  P_\mu P^\mu = P^+ P^-  \! -  \mbf{P}_\perp^2$~\cite{{Brodsky:2014yha}} 
\begin{align} \lb{P2M2}
P^2 \vert  \psi(P) \rangle =  M^2 \vert  \psi(P) \rangle.
\end{align}
The simple structure of the LF vacuum allows for a quantum-mechanical probabilistic interpretation of hadron states in terms of the eigenfunctions of the LF Hamiltonian equation \req{P2M2} in a constituent particle basis, $\vert \psi \rangle = \sum_n \psi_n  \vert n \rangle$, written in terms of the quark and gluon degrees of freedom in the Fock expansion.  

In practice, solving the actual eigenvalue problem \req{P2M2} is a formidable computational task  for a non-abelian quantum field theory beyond 1 + 1 dimensions, and particularly in three and four-dimensional space-time, with an unbound particle number  with arbitrary momenta and helicities. Consequently, alternative methods and approximations  are necessary to tackle the relativistic bound-states in the strong-coupling regime of QCD.

\subsection{QCD (1 + 1)}

The  ’t Hooft model~\cite{tHooft:1974pnl} in one-space and one-time dimensions, constitutes the first example of a semiclassical Hamiltonian wave equation derived from first principles QCD in light-front quantization~\cite{Dirac:1949cp}. This equation is exact in the large-$N$ limit and leads to the computation of a meson spectrum and light front wave functions in terms of the constituent quark and antiquark, while incorporating chiral symmetry breaking (CSB) and confinement. 
 
In QCD (1 + 1) gluons are not dynamical, there are no gluon self-couplings, and quarks have chirality but no spin. The coupling $g$ has dimension of mass and it is a confining gauge theory for any value of the coupling.  We can express the QCD Lagrangian \req{L} in 1 + 1 dimensions, with LF coordinates $x^+ = x^0 + x^3$ and $x^- = x^0 - x^3$, in the $A^+ = 0$  gauge  in terms of the fields $\psi_{\pm} \equiv \psi_{R,L}$  and $A^-$. The LF constraint equations imply that there is only one independent degree of freedom, $\psi_+$. The hadron 2-momentum  generator $P =  (P^+, P^-)$,  $P^\pm = P^0 \pm P^3$, is then expressed in terms of the field $\psi_+$~\cite{Hornbostel:1988ne, Hornbostel:1988fb, Eller:1986nt} with
\begin{align}  \lb{Pm11}
P^-  =  {\half} \int   dx^-  \Big( \psi_+^\dagger  \frac{ m^2 }{ i \partial^+}  \psi_+
 +  g^2 j^{+a}  \frac{1}{ (i \partial^+)^2}   j^{+a}\Big),
  \end{align} 
for the LF Hamiltonian where $ j^{+ a} = \psi_+^\dagger  T^a \psi_+$.  From the inverse derivative in the interaction term in \req{Pm11}  (the term with the coupling) there follows the potential $V$
\begin{align}
V = -  g^2 \int d x^-  d y^-  j^{+ a}(x^-) \left \vert x^- -  y^- \right \vert j^{+ a}(y^-).
 \end{align}

The pion mass spectrum can be computed from the LF eigenvalue equation  \req{P2M2} for QCD(1+1), namely  $P^+ P^- \vert  \chi(P^+) \rangle =  M_\pi^2 \vert  \chi(P^+)\rangle$.  For the $q \bar q$ valence state  it leads to~\cite{Hornbostel:1988ne, Eller:1986nt}   \begin{align} \lb{tHE}
 \Big( \frac{m_q^2}{x} +  \frac{m_{\bar q}^2}{1-x}\Big) \chi(x) + \frac {\lambda_N}{\pi}   P \! \int_0^1 \! dx'  \, \frac{\chi(x) - \chi(x')}{(x - x')^2} = M_\pi^2  \, \chi(x),
\end{align}
the $`$t Hooft equation~\cite{tHooft:1974pnl} with effective coupling $\lambda_N =  g^2 \left(N^2 - 1\right) / 2 N$, where $x$ is the longitudinal momentum fraction of the $q \bar q$ state. Cancellation of singularities at $x = \epsilon$ and $x =1 - \epsilon$ for the approximate solution 
$ \chi(x) \sim x^{\beta_q} (1-x)^{\beta_{\bar q}}$  in Eq.~\req{tHE} leads, for $m_q^2/\pi \lambda_N \ll 1$,  to
$\beta_q = \left( 3 m_q^2/\pi \lambda_N \right)^{{1}/{2}}$ and
\begin{align}
M_\pi^2  =   \sqrt{\frac{\pi \lambda_N}{ 3}} \, (m_q + m_{\bar q}) +  \mathcal{O} \! \left( (m_q \! + m_{\bar q})^2 \right).
\end{align}

In QCD (1+1),  the value of the CSB  ``condensate’’ 
$\langle \bar \psi \psi \rangle =  - f_\pi^2 \sqrt{\pi \lambda_N/ 3}$
and the strength of the linear confinement depend on the value of the coupling $g$ in the  QCD Lagrangian, and are not emerging properties\footnote{The glueball spectrum has been computed in a large-$N$ model of QCD in 1 + 1 dimensions~\cite{Demeterfi:1993rs}.}.

\subsection{\lb{3p1}QCD (3 + 1)}

Our purpose is to find a semiclassical approximation to strongly coupled QCD  to compute hadronic bound states and other hadronic properties. To this end, it is necessary to reduce the multiparticle eigenvalue problem of the LF Hamiltonian (\ref{P2M2}) to an effective relativistic light-front wave equation, instead of diagonalizing the full Hamiltonian. The central problem then becomes the derivation of the effective confining interaction, which acts only on the valence sector of the theory and has, by definition, the same eigenvalue spectrum as the initial Hamiltonian problem. 

In 3 + 1 dimensions we also start with the QCD Lagrangian in \req{L} and assume that, to a first semiclassical approximation, gluons with small virtualities are non-dynamical and incorporated in the confinement potential. This approximation entails an important simplification of the full LF Hamiltonian $P^-$, which we express in terms of the dynamical quark field $\psi_+$, $\psi_\pm = \Lambda_\pm \psi$ ,  $\Lambda_\pm = \gamma^0 \gamma^\pm$ in the $A^+ = 0$ gauge~\cite{deTeramond:2008ht}
\begin{align}  \lb{Pm31}
P^-  =  {\half} \int \! dx^- d^2 \mbf{x}_\perp \bar \psi_+ 
\frac{ \left( i \mbf{\nabla}_{\! \perp} \right)^2 + m^2 }{ i \partial^+}  \psi_+
 + ~ {\rm  interactions}.
  \end{align}

For a $q \bar q$ bound state we factor out the longitudinal $X(x)$ and orbital $e^{i L \theta}$  dependence from  the LF wave function $\psi$, 
\begin{align}
\psi(x,\zeta, \theta) = e^{i L \theta} X(x)  \frac{\phi(\zeta)}{\sqrt{2 \pi \zeta}},
\end{align}
with longitudinal momentum fraction $x$. The variable  $\zeta^2 = x(1-x)  \mbf{b}^2_\perp$ is the invariant transverse separation between two quarks, with $\mbf{b}_\perp$, the relative impact variable, conjugate to the relative transverse momentum $\mbf{k}_\perp$.  In the  ultra-relativistic zero-quark mass limit $m_q  \to  0$ the longitudinal modes $X(x)$ decouple and we find
\begin{align}
M^2  =  \int \! d\zeta \, \phi^*(\zeta) \sqrt{\zeta}
\left( -\frac{d^2}{d\zeta^2} -\frac{1}{\zeta} \frac{d}{d\zeta}
+ \frac{L^2}{\zeta^2}\right)
\frac{\phi(\zeta)}{\sqrt{\zeta}} 
+  \int \! d\zeta \, \phi^*(\zeta) \,U(\zeta) \, \phi(\zeta) ,
\end{align}
where the  effective potential $U$ incorporates all interactions,  including those from higher Fock states. The Lorentz invariant LF equation~\req{P2M2},  $P_\mu P^\mu  \vert \psi \rangle = M^2 \vert \psi \rangle$,
 becomes a LF wave equation for $\phi$~\cite{deTeramond:2008ht}
\begin{align} \lb{LFWE}
\left(-\frac{d^2}{d\zeta^2} 
- \frac{1 - 4L^2}{4\zeta^2}+ U(\zeta) \right)  \phi(\zeta) = M^2 \phi(\zeta),
\end{align}
where the critical value of the LF orbital angular momentum $L = 0$ corresponds to the lowest possible stable solution, the ground state of the light-front Hamiltonian. Eq.~\req{LFWE} is relativistic and frame-independent; It has a similar structure to wave equations in AdS provided that one identifies $\zeta = z$, the holographic variable~\cite{deTeramond:2008ht}.

\section{Higher spin wave equations in AdS}

Anti-de Sitter AdS$_{d+1}$ is the maximally symmetric $d+1$ space with negative constant curvature and a $d$-dimensional flat space boundary, Minkowski spacetime.  In Poincar\'e  coordinates   $x^M = \left(x^0, x^1, \cdots , x^d, z\right)$, where the asymptotic border  of AdS space  is given by $z = 0$, the line element  is
\begin{align} \label{AdSm}
ds^2  &= g_{MN}dx^M dx^N  \nn \\
&= \frac{R^2}{z^2} \left(\eta_{\mu \nu} dx^\mu dx^\nu - dz^2\right),
\end{align}
where $\eta_{\mu \nu}$ is the usual  Minkowski metric in $d$ dimensions and $R$ is the AdS radius. Five-dimensional anti-de Sitter space, AdS$_5$, has 15 isometries which induce in the Minkowski spacetime boundary the symmetry under the conformal group with 15 generators in four dimensions. This conformal symmetry implies that there can be no scale in the boundary theory and therefore no discrete spectrum.

\begin{figure}[ht]
\begin{center}
        \includegraphics[width=6.8cm]{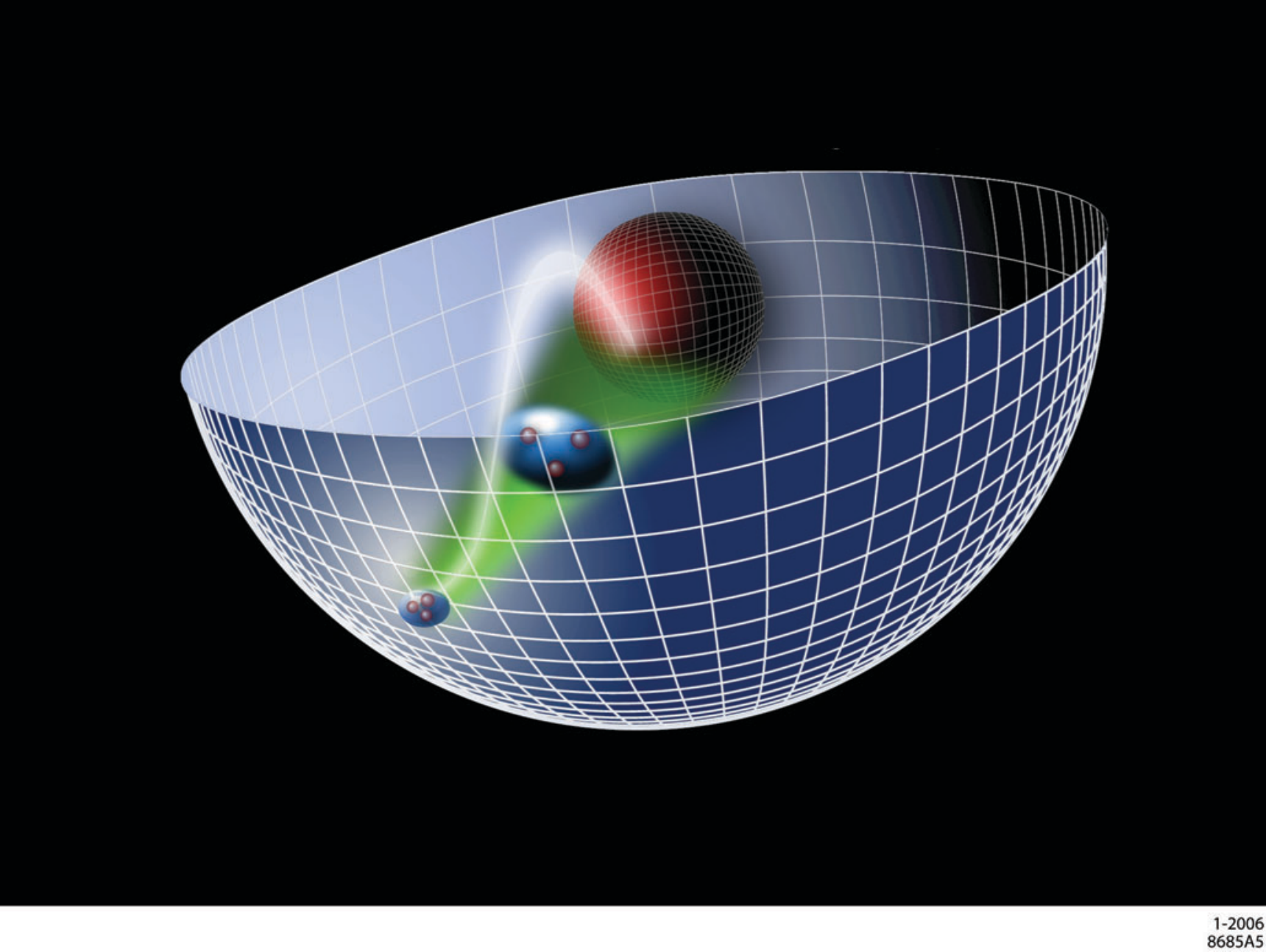}
        \end{center}
        \caption{\small Different values of the AdS radius $z$ correspond to different scales at which the proton is examined. Events at short distances in the ultraviolet happen near the four-dimensional AdS boundary (large circumference),  where the proton size shrinks as perceived by an observer in Minkowski space due to the negative curvature of AdS. The inner sphere represents large distance infrared events where AdS is modified to model confinement. \lb{fig:hologproton}}
\end{figure}

The variable $z$ acts like a scaling variable in Minkowski space: different values of $z$ correspond to different energy scales at which a measurement is made.  Short spacetime intervals map to the boundary in AdS space-time  near $z=0$. This corresponds to the ultraviolet (UV) region of AdS space, the conformal boundary.   On the other hand, a large  four-dimensional object of confinement dimensions  of hadronic size maps to the large infrared~(IR) region of AdS. Thus,  in order to incorporate  confinement and discrete normalizable modes (Fig. \ref{fig:hologproton}) in the gravity dual, the conformal invariance must be broken by modifying  AdS space  in the IR large $z$  region,  by introducing, for example,  a sharp cut-off at the IR border,  as in the ``hard-wall’’  model of Ref.~\cite{Polchinski:2001tt},  or  by using a  ``soft-wall’’ model~\cite{Karch:2006pv}  to reproduce the observed linearity of Regge trajectories.

\subsection{Integer-spin wave equations}

The holographic embedding of the semiclassical LF bound-state wave equation  in AdS space allows us to extended \req{LFWE} to arbitrary integer spin $J$, by studying the propagation of spin-$J$ modes in AdS space~\cite{deTeramond:2008ht, deTeramond:2013it}. To examine the specific mapping of the wave equations in AdS to LF physics, we start with the AdS action for a tensor-$J$ field $\Phi_J = \Phi_{N_1 \dots N_J}$ in the presence of a dilaton profile $\varphi(z)$ responsible for the confinement dynamics
  \begin{align} \lb{SAdS}
  S = \int d^dx \,  dz \sqrt{g} \, e^{\varphi(z)}  \left(D_M \Phi_J D^M \Phi_J  - \mu^2 \Phi_J^2\right),
  \end{align}
  where $g$ is the determinant of the metric tensor $g_{MN}$,   $d$ is the number of transverse coordinates, and $D_M$ is the covariant derivative which includes the affine connection. The  variation of the AdS action leads to the wave equation
  \begin{align} \lb{AdSWEJ}
      \left[
   -  \frac{ z^{d-1- 2J}}{e^{\varphi(z)}}   \partial_z \Big(\frac{e^{\varphi(z)}}{z^{d-1-2J}} \partial_z   \Big)
  +  \frac{(\mu\,R )^2}{z^2}  \right]  \Phi_J(z) = M^2 \Phi_J(z), 
  \end{align}
after a redefinition of the AdS mass $\mu$,  plus kinematical constraints to eliminate lower spin from the symmetric tensor $\Phi_{N_1 \dots N_J}$~\cite{deTeramond:2013it}. 

By substituting $\Phi_J(z) = z^{(d-1)/2 -J} e^{- \varphi(z)/2} \, \phi_J(z)$ in \req{AdSWEJ}, we find  the semiclassical light-front wave equation~\req{LFWE} with 
  \begin{align} \lb{Uvarphi}
      U_J(\zeta) = \frac{1}{2} \varphi''(\zeta) + \frac{1}{4} \varphi'(\zeta)^2 + \frac{2J - 3}{2 \zeta} \varphi’(\zeta),
  \end{align} 
for $d = 4$ as long as $\zeta = z$. The precise mapping allows us to write the LF confinement potential $U$ in terms of the dilaton profile which modifies the IR region of AdS space to incorporate confinement~\cite{Brodsky:2014yha}, while keeping the theory conformal invariant in the ultraviolet boundary of AdS for $z \to 0$. The separation of kinematic and dynamic components, allows us to determine the mass function in the AdS action in terms of physical kinematic quantities with the AdS mass-radius $(\mu R)^2 =  L^2 - (2-J)^2$~\cite{deTeramond:2008ht, deTeramond:2013it}.

\subsection{Half-integer-spin wave equations}

A similar derivation follows from the Rarita-Schwinger action for a spinor field $\Psi_J \equiv \Psi_{N_1 \dots N_{J - 1/2}}$ in AdS for  half-integral spin $J$~\cite{deTeramond:2013it}. In this case, however, the dilaton term does not lead to an interaction~\cite{Kirsch:2006he},  and an effective Yukawa-type coupling to a potential $V$ in the action has to be introduced instead~\cite{deTeramond:2013it, Abidin:2009hr, Gutsche:2011vb}:
  \begin{align} \lb{SAdSs}
  S = \int \, d^4x \,  dz \sqrt{g}  \, \bar \Psi_J  \left(i \Gamma^A e_A^M D_M  - \mu + \frac{z}{R} V(z) \right) \Psi_J,
  \end{align}
 where $e^M_A$ is the vielbein and the covariant derivative $D_M$ on a spinor field includes the affine connection and the spin connection. The tangent space Dirac matrices, $\Gamma^A$, obey the usual anticommutation relations $\{\Gamma^A, \Gamma^B\} = 2 \eta^{A B}$.
Factoring out the four-dimensional plane-wave and spinor dependence, 
$\Psi_J(x, z)_\pm   = e^{ i P \cdot x}  z^{(d+1)/2 - J} \,   \psi_\pm(z) \, u^\pm_{\nu_1 \cdots \nu_{J-1/2}}({P})$,
we find from \req{SAdSs} the coupled linear differential equations for the chiral components $\psi_\pm$  
\begin{align} 
- \frac{d}{d z} \psi_-  - \frac{\nu+\half}{z}\psi_-  -  V(z) \psi_-&= M \psi_+ , \lb{psip} \\
 \frac{d}{d z} \psi_+ - \frac{\nu+\half}{z}\psi_+  - V(z) \psi_+ &= M \psi_-  ,  \lb{psim}
\end{align}
where  $\vert\mu R \vert = \nu + \half$. 

Eqs. \req{psip} and \req{psim} are equivalent to the second order equations~\cite{deTeramond:2013it}
\begin{align} 
\left(-\frac{d^2}{d\zeta^2}
- \frac{1 - 4 L^2}{4\zeta^2} + U^+(\zeta) \right) \psi_+\! & =  M^2 \psi_+ \lb{psi1} , \\
\left(-\frac{d^2}{d\zeta^2} 
- \frac{1 - 4(L + 1)^2}{4\zeta^2} + U^-(\zeta) \right) \psi_- &=M^2 \psi_-   \lb{psi2} ,
\end{align}
with $\zeta = z$,   $\nu = L$, and
\begin{align} \lb{UV}
U^\pm(\zeta) = V^2(\zeta) \pm V'(\zeta) + \frac{1 + 2 L}{\zeta} \, V(\zeta) , 
\end{align}
with equal probability $\int d\zeta \, \psi_+^2(\zeta)^2 = \int d\zeta  \, \psi_-^2(\zeta)$. The semiclassical LF wave equations for  $\psi_+$ and $\psi_-$ correspond to the LF orbital angular momentum $L$ and $L + 1$: their structure is compatible with the observed degeneracy of baryon states with the same orbital quantum number $L$, namely the absence of spin-orbit coupling.


 \section{Superconformal algebraic structure and emergence of a mass scale}

The precise mapping of the semiclassical light-front Hamiltonian equations to the wave equations in AdS space gives important insights into the nonperturbative structure of bound state equations in QCD for arbitrary spin, but it does not answer the question of how the effective confinement dynamics is actually determined, and how it can be related to the symmetries of QCD itself. An important clue, however, comes from the realization that the potential $V(\zeta)$ in Eq.~\req{UV} plays the role of the superpotential  in supersymmetric (SUSY) quantum mechanics (QM)~\cite{Witten:1981nf}.  In fact, the idea to apply an effective  supersymmetry to hadron physics is certainly not new~\cite{Miyazawa:1966mfa, Catto:1984wi, Lichtenberg:1999sc}, but failed to account for the special role of the pion.  In contrast in the HLFQCD approach, as we shall discuss below, the zero-energy normalizable eigenstate of the superconformal quantum mechanical equations is identified with the pion and has no baryonic supersymmetric partner, a pattern which is observed across all the particle families~\cite{ParticleDataGroup:2022pth}.

Supersymmetric QM is based on a graded Lie algebra consisting of two anticommuting supercharges $Q$ and $Q^\dagger$,  $\{Q,Q\} =  \{Q^\dagger, Q^\dagger\} = 0$, which commute with the Hamiltonian $H = \tfrac{1}{2} \{Q, Q^\dagger\}$, $[Q, H]  = [Q^\dagger, H] = 0$.  If the state $\vert E \rangle$ is an eigenstate with energy $E$, $H\vert E \rangle = E \vert E \rangle$, then, it follows from the commutation relations that the state $Q^\dagger \vert E \rangle$  is degenerate with the state $\vert E \rangle$ for $E \ne 0$, but for  $E = 0$  we have  $Q^\dagger \vert E=0 \rangle= 0$. This is a key result for deriving the supermultiplet structure and the pattern of the hadron spectrum since the zero mode has no supersymmetric partner~\cite{Witten:1981nf}.

Following Ref.~\cite{Fubini:1984hf} we consider the scale-deformed supercharge operator  $R_\lambda = Q~+~\lambda S$ with $K = \tfrac{1}{2} \{S, S^\dagger\}$, the generator of special conformal transformations. The generator $R_\lambda$ is also nilpotent, $\{R_\lambda, R_\lambda\} =  \{R_\lambda^\dagger, R_\lambda^\dagger\} = 0$, and gives rise to a new scale-dependent Hamiltonian $G$,  $G = \tfrac{1}{2} \{R_\lambda, R_\lambda^\dagger\}$, which also closes under the graded algebra, $[R_\lambda, G]  = [R_\lambda^\dagger, G] = 0$.  The new supercharge $ R_\lambda$ has the matrix representation 
 \begin{align} \lb{Rex} 
 R_\lambda = 
 \left(\begin{array}{cc} 0 & r_\lambda \\ 0 & 0 \end{array}\right), \quad
R_\lambda^\dagger=\left(\begin{array}{cc} 0 & 0\\r^\dagger _\lambda&0\end{array}\right),
\end{align}
with $r_\lambda = - \partial_x+\frac{f}{x}+\lambda x,\; r^\dagger _\lambda  = \partial_x+\frac{f}{x}+\lambda x$. The parameter $f$ is dimensionless and $\lambda$ has the dimension of  [$M^2$]; Therefore, a mass scale is introduced in the Hamiltonian without leaving the conformal group. The Hamiltonian equation $G\vert E \rangle = E \vert E \rangle$ leads to the wave equations
\begin{align} 
 \left(\! - \frac{d^2}{d x^2}  - \frac{1-  4 (f + \half)^2}{4 x^2} + \lambda ^2 \,x^2+ 2 \lambda \left( f  - \half\right) \right)  \! \phi_+ 
 = E  \phi_+,   \lb{phi1} ~~~~~~ \\
 \left(\! - \frac{d^2}{d x^2}  - \frac{1- 4 (f - \half)^2}{4 x^2} + \lambda^2 \, x^2 +   2 \lambda \left( f  + \half\right) \right) \! \phi_-  
 =  E  \phi_- ,  \lb{phi2} ~~~~~~
 \end{align}
which have the same structure as the Euler-Lagrange equations obtained from the AdS holographic embedding of the LF Hamiltonian equations, but here,  the form of the LF confinement potential, $\lambda^2 x^2$, as well as the constant terms in the potential are completely fixed by the superconformal symmetry~\cite{deTeramond:2014asa, Dosch:2015nwa}.

\subsection{Light-front mapping and baryons}
 
\begin{figure}[ht]
\begin{center}
\includegraphics[width=7.4cm]{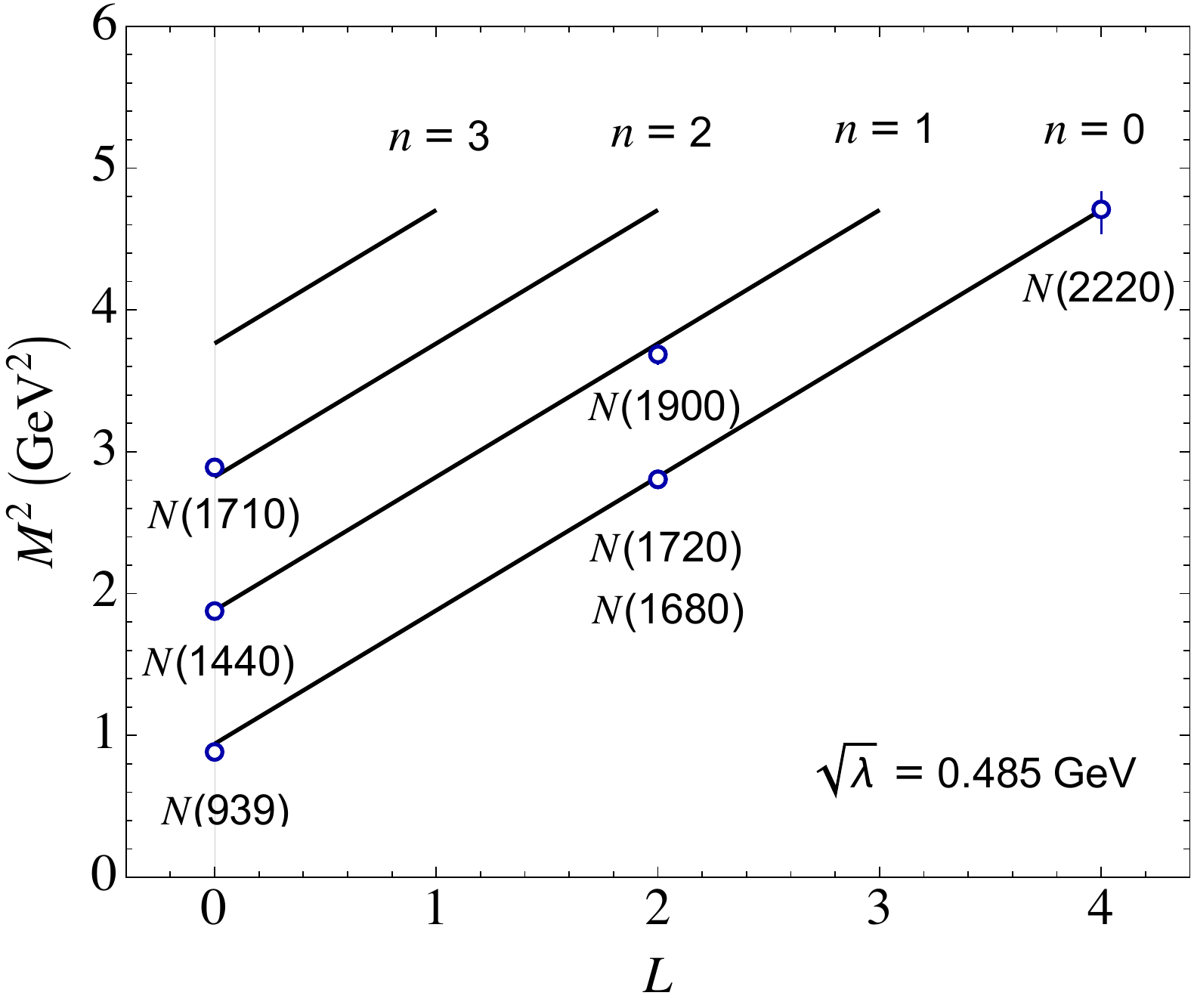} \hspace{20pt}
\includegraphics[width=7.4cm]{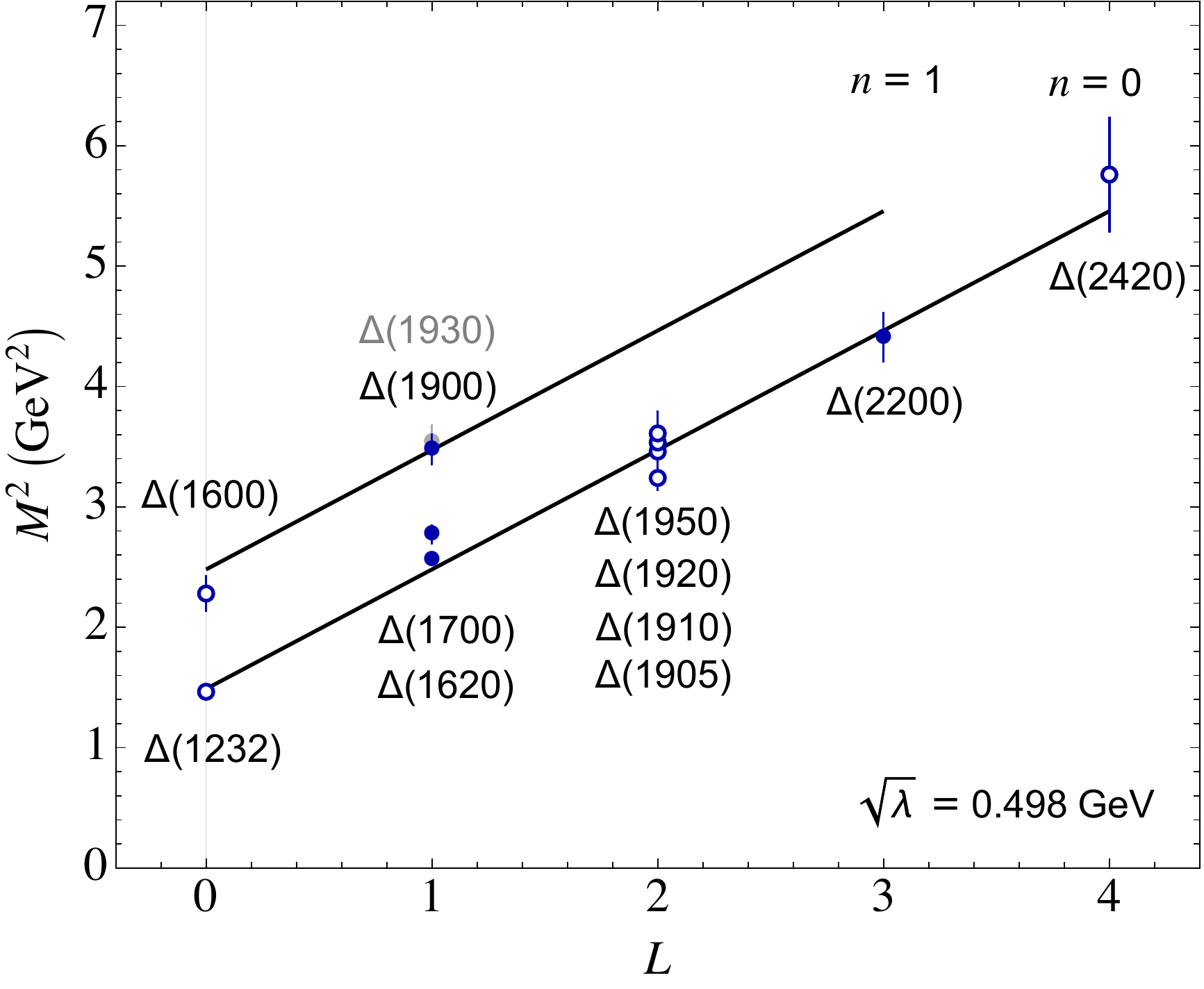}
\end{center}
\caption{\lb{fig:nucleon-delta} \small Model predictions for the orbital and radial positive-parity nucleons (left) and positive and negative  parity $\Delta$ families  (right) compared with the data from Ref.~\cite{ParticleDataGroup:2022pth}. The values of $\sqrt{\lambda}$ are $\sqrt{\lambda} = 0.485$ GeV for nucleons and $\sqrt{\lambda} = 0.498$ GeV for the deltas.}
\end{figure}

Upon mapping \req{phi1} and \req{phi2} to the semiclassical LF wave equations \req{psi1} and \req{psi2} using the substitutions
$~x \mapsto \zeta, ~ E \mapsto M^2,  ~ f  \mapsto   L + \half, ~
 \phi_+ \mapsto \psi_- ~ {\rm and}~  \phi_- \,  \mapsto \psi_+ $, we find the result 
 \begin{align}
 U^+ &= \lambda^2 \zeta^2 + 2 \lambda(L + 1), \\
 U^- &=  \lambda^2 \zeta^2 + 2 \lambda L,
 \end{align}
 for the confinement potential of baryons~\cite{deTeramond:2014asa}. The solution of the LF wave equations for this potential  gives  the eigenfunctions  
 \begin{align}
 \psi_+(\zeta)  & \sim \zeta^{\frac{1}{2} + L} e^{-\lambda \zeta^2/2}
  L_n^L(\lambda \zeta^2), \\
\psi_-(\zeta) & \sim \zeta^{\frac{3}{2} + L} e^{-\lambda \zeta^2/2} L_n^{L+1}(\lambda \zeta^2) ,
\end{align}
 with eigenvalues  
 \begin{align}
 M^2 = 4 \lambda (n + L + 1).
 \end{align}  
 The polynomials $L_n^L(x)$ are associated Laguerre polynomials, where the radial quantum number $n$ counts the number of nodes in the wave function. We compare in Fig. \ref{fig:nucleon-delta} the model predictions with the measured values for the positive parity nucleons~\cite{ParticleDataGroup:2022pth} for $\sqrt{\lambda} = 0.485$ GeV.

 \subsection{\lb{scMB}Superconformal meson-baryon symmetry}

Superconformal quantum mechanics also leads to a connection between mesons and baryons~\cite{Dosch:2015nwa} underlying the $SU(3)_C$ representation properties, since a diquark cluster can be in the same color representation as an antiquark, namely $ \bf{\bar 3} \in {\bf 3} \times {\bf 3}$. The specific connection follows from the substitution $x  \mapsto \zeta, \  E \mapsto M^2,  \
\lambda \mapsto \lambda_B=\lambda_M, \ f  \mapsto L_M-\half = L_B + \half,\ \phi_+ ~\mapsto ~\phi_M \ {\rm and} \ \phi_- \mapsto \phi_B$  in the superconformal equations \req{phi1} and \req{phi2}. We find the LF meson (M) -- baryon (B) bound-state equations
\begin{align} \lb{M}
 \left(-\frac{d^2}{d\zeta^2} - \frac{1- 4 L_M^2}{4 \zeta^2} +  U_M \right)\phi_M  &= M^2 \, \phi_M  , \\ \lb{B}
\left(-\frac{d^2}{d\zeta^2} -  \frac{1- 4 L_B^2}{4 \zeta^2} + U_B   \right)\phi_B &= M^2 \, \phi_B , 
\end{align}
with the confinement potentials 
\begin{align}
U_M &=  \lambda_M^2\, \zeta^2 + 2 \lambda_M (L_M - 1) ,  \lb{UM} \\  
U_B  &= \lambda_B^2\, \zeta^2 +  2 \lambda_B (L_B +1).    \lb{UB}
\end{align}

The superconformal structure imposes the condition  $\lambda = \lambda_M = \lambda_B$  and the remarkable relation $L_M = L_B + 1$, where $L_M$ is the  LF angular momentum between the quark and antiquark in the meson, and $L_B$ between the active quark and spectator cluster in the baryon. Likewise, the equality of the Regge slopes embodies the equivalence of the ${\bf 3} -{\bf\bar 3}$ color interaction  in the $q \bar q$ meson with the ${\bf 3} - {\bf\bar 3}$ interaction between the quark and diquark cluster in the baryon. The mass spectrum from \req{M} and \req{B} is 
\begin{align} \lb{MNspec}
M^2_M &= 4 \lambda (n+ L_M ) , \\
M^2_B &= 4 \lambda (n+ L_B+1).
\end{align}
The pion has a special role as the unique state of zero mass and, since $L_M = 0$, it has not a baryon partner.  

\begin{figure} 
\begin{center}
\includegraphics[width=7.4cm]{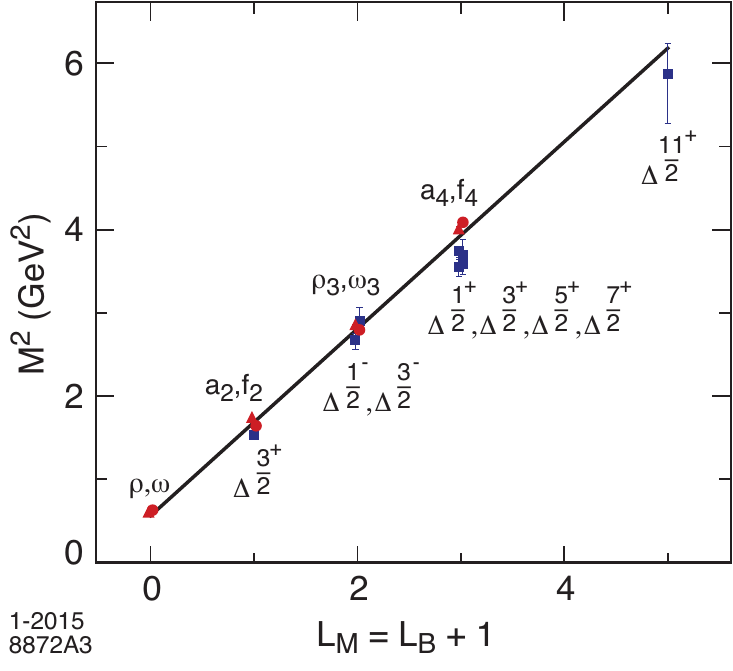}
\end{center}
\caption{\lb{fig:rho-delta} \small Supersymmetric vector meson and $\Delta$  partners from Ref.~\cite{Dosch:2015nwa}.  The experimental values of $M^2$ from Ref.~\cite{ParticleDataGroup:2022pth}  are plotted vs $L_M = L_B+1$ for  $\sqrt \lambda \simeq 0.5$ GeV. The $\rho$ and $\omega$ mesons have no baryonic partner, since it would imply a negative value of $L_B$.}
\end{figure}

\subsection{Spin interaction and diquark clusters}

Embedding the LF  equations in AdS space allows us to extend the superconformal Hamiltonian to include the spin-spin interaction, a problem not well defined in the chiral limit by standard procedures. The dilaton profile $\varphi(z)$ in the AdS action \req{SAdS} can be determined from the  superconformal algebra by integrating  Eq. \req{Uvarphi} for the effective potential $U$ \req{UM} in Eq.~\req{M}, $U = \lambda^2 z^2 + 2 \lambda (L - 1)$. One obtains the result $\varphi(z) = \lambda z^2$ which is uniquely determined, provided that it depends only on the modification of AdS space~\cite{Dosch:2016zdv}. Since the dilaton profile $\varphi(z) = \lambda z^2$ is valid for arbitrary $J$, it leads to the additional term  $2 \lambda \Scal$ in the LF Hamiltonian for mesons and baryons,   $G = \tfrac{1}{2} \{R_\lambda, R_\lambda^\dagger\} + 2 \lambda \Scal$, which maintains the meson-baryon supersymmetry~\cite{Brodsky:2016yod}. The spin $\Scal = 0, 1$, is the total internal spin of the meson, or the spin of the diquark cluster of the baryon partner. The effect of the spin term is an overall shift of the quadratic mass
\begin{align}
M^2_M &= 4 \lambda (n+ L_M )  + 2 \lambda \Scal  \lb{M2S}, \\
M^2_B &= 4 \lambda (n+ L_B+1) + 2 \lambda \Scal  \lb{M2},
\end{align}
as depicted in Fig.~\ref{fig:rho-delta} for the spectra of the $\rho$ mesons and $\Delta$ baryons~\cite{Dosch:2015nwa}. For the $\Delta$ baryons the total internal spin $S$ is related to the diquark cluster spin $\Scal$ by $S = \Scal  + \frac{1}{2} (-1)^L$, and therefore, positive and negative $\Delta$ baryons have the same diquark spin, $\Scal = 1$.  As a result, all the $\Delta$ baryons lie, for a given $n$, on the same Regge trajectory as shown in  Fig.~\ref{fig:nucleon-delta}. For negative parity nucleons  both  $\Scal = 0$  and  $\Scal = 1$ are possible, but their systematics is less well understood as compared with the $\Delta$ baryons and positive parity nucleons.

\subsection{Inclusion of quark masses}

In the usual formulation of bottom-up holographic models, one identifies quark masses and chiral condensates as coefficients of a scalar background field $X_0(z)$ in AdS space~\cite{Erlich:2005qh, DaRold:2005mxj}. A heuristic way to take into account the occurrence of quark mass terms in the HLFQCD approach,  is to include the quark mass dependence in the invariant mass which controls the off-shell dependence of the LF wave function.  This substitution leads, upon exponentiation, to  a natural factorization of the  transverse, $\phi(\zeta)$,  and the longitudinal, $\chi(x)$, wave functions. For hadrons with quark masses  $m_i$ one finds~\cite{Brodsky:2014yha}
\begin{align}   \lb{IMWF} 
\chi(x) \sim 
 \exp \Big( \! - \frac{1}{2 \lambda} {\small \sum_i} \frac{m_i^2}{x_i}  \Big),
\end{align}
but it is not a unique prescription~\cite{deTeramond:2021yyi, Li:2022izo}.   This approach  has been consistently applied to the radial and orbital excitation spectra of the light  $\pi, \rho, K, K^*$ and $\phi$ meson families,  as well as to the $N, \Delta, \Lambda, \Sigma, \Sigma^*, \Xi$ and $\Xi^*$ in the baryon sector,  giving the value  $\sqrt{\lambda} = 0.523 \pm 0.024$ GeV~\cite{Brodsky:2016yod}. The comparison of the predicted $K^*$ and $\Sigma^*$  trajectories  with experiment in Fig.~\ref{KSigma} is a clear example of the validity of the supersymmetric meson-baryon connection including light quark masses.

\begin{figure}[ht]
\begin{center}
\includegraphics[width=8.4cm]{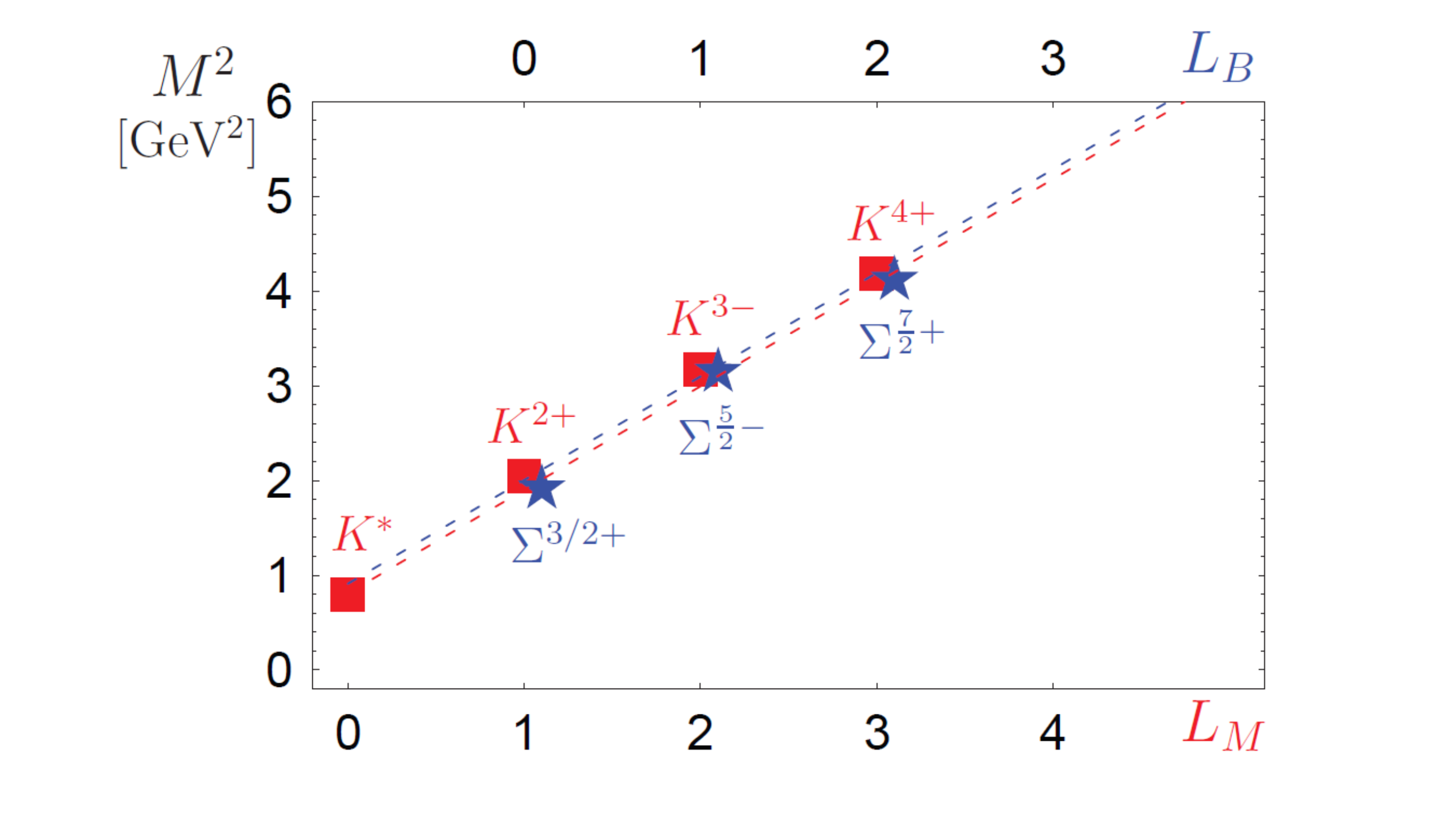} 
\end{center}
\caption{ \lb{KSigma}  \small Supersymetric $K^*$ and $\Sigma^*$ trajectories  in HLFQCD from Ref.~\cite{Dosch:2020hqm} with $\sqrt\lambda = 0.51$  GeV.}
\end{figure}

For heavy quarks the mass breaking effects are large. The underlying hadronic supersymmetry, however,  is still compatible with the holographic approach and gives remarkable connections across the entire spectrum of light and heavy-light hadrons~\cite{Dosch:2016zdv, Dosch:2015bca}.  In particular, the lowest mass meson defining the $K, K^*, \eta', \phi, D, D^*, D_s, B, B^*,B_s$ and $B^*_s$ families has no baryon partner, conforming to the SUSY mechanism found for the light hadrons, and depicted in Fig.~\ref{fig:rho-delta}. The heavy quark symmetry~\cite{Shuryak:1981fza, Isgur:1991wq} predicts a dependence of the holographic mass scale $\lambda$ on the quark mass~\cite{Dosch:2016zdv, Gutsche:2012ez}, $\sqrt{\lambda_M} = C \sqrt{M_M}$, where $M_M$ is the mass of the lowest state containing one or two heavy quarks with $\sqrt{C}= 0.49   \pm 0.02 \, {\rm GeV}^{1/2}$~\cite{Dosch:2020hqm, Nielsen:2018ytt}.

 \subsection{Completing the supersymmetric hadron multiplet}

\begin{figure}[]
\begin{center}
\includegraphics[width=7.2cm]{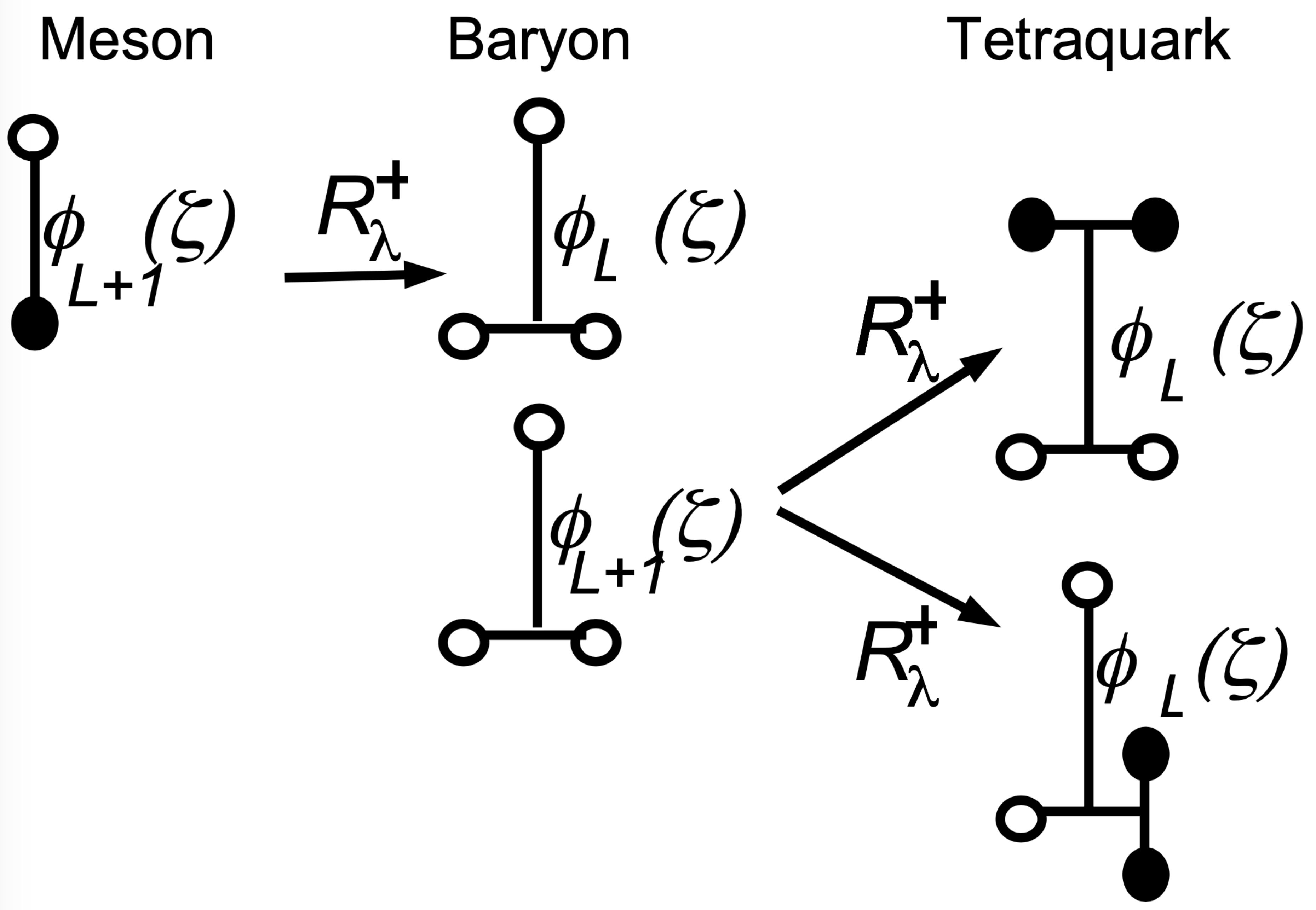} 
\end{center}
\caption{\lb{fig:MBTplet} \small The meson-baryon-tetraquark supersymmetric 4-plet $\{\phi_M, \phi_B^+, \phi_B^-, \phi_T\}$ follows from the two step action of the supercharge operator $R^\dagger_\lambda$:   ${\bf \bar 3} \to {\bf 3} \times {\bf 3}$ on the pion, followed by  $ {\bf 3} \to {\bf \bar 3} \times {\bf \bar 3}$ on the negative chirality component of the nucleon.}
\end{figure}

Besides the mesons and the baryons, the supersymmetric multiplet $\Phi = \{\phi_M, \phi_B^+, \phi_B^-, \phi_T\}$  contains a further bosonic partner, a tetraquark, which, as illustrated in Fig.~\ref{fig:MBTplet},  follows from the action of the supercharge operator $R^\dagger_\lambda$ on the negative-chirality component of a baryon~\cite{Brodsky:2016yod}.  A clear example is the SUSY positive parity $J^P$-multiplet  $2^+,   \frac{3}{2}^+,   1^+$  of states  $f_2(1270),  ~\Delta(1232),  ~a_1(1260)$ where the $a_1$ is interpreted as a tetraquark.

\begin{table}[ht]
\centering
\caption{\lb{pred} \small Predicted masses for double heavy bosons from Ref.~\cite{Dosch:2020hqm}.  Exotics which are predicted to be stable under strong interactions are marked by $^{(!)}$.}
\vspace{2mm}
\begin{tabular}{c|c|c|c|c}  
\hline \hline 
quark & $J^P$&predicted&strong&threshold \vspace{-1mm}\\
 content&&Mass [MeV]&decay&[MeV]\\
\hline
$cq\overline{cq}$&$ 0^+ $& 3660& $\eta_c\pi\pi $ &3270\\
$cc\overline{qq} ^{(!)} $&$ 1^+ $& 3870& $D^*D $ &3880\\
\hline
$bq\overline{bq}$&$ 0^+ $& 10020& $\eta_b\pi\pi $ &9680\\
$bb\overline{qq} ^{(!)} $&$ 1^+ $& 10230& $B^*B $ &10800\\
\hline
$bc\overline{qq} ^{(!)} $&$ 0^+ $& 6810& $BD $ &7150\\
\hline \hline
\end{tabular}
\end{table}

Unfortunately, it is difficult to disentangle conventional hadronic quark states from exotic ones and, therefore, no clear-cut identification of tetraquarks for light hadrons, or hadrons with hidden charm or beauty, has been found~\cite{Brodsky:2016yod,  Nielsen:2018ytt, Nielsen:2018uyn}. The situation is, however, more favorable for tetraquarks with open charm and beauty which may be stable under strong interactions and therefore easily identified~\cite{Karliner:2017qjm}. In Table~\ref{pred}, the computed masses from Ref.~\cite{Dosch:2020hqm} are presented.  Our prediction~\cite{Dosch:2020hqm} for a doubly charmed stable boson $T_{cc}$  with a mass of 3870 MeV (second row) has been observed at LHCb a year later at 3875 MeV~\cite{LHCb:2021vvq}, and it is a member of the positive parity $J^P$-multiplet  $2^+,   \frac{3}{2}^+,   1^+$  of states  $\chi_{c2}(3565), ~ \Xi_{cc}(3770), ~ T_{cc}(3875)$. The possible occurrence of stable doubly beautiful tetraquarks and those with charm and beauty is well founded~\cite{Karliner:2017qjm}.

 \section{Summary and outlook}

Holographic light front QCD is a nonperturbative analytic approach to hadron physics. It originates in the precise mapping between the hadron form factors in AdS space and physical spacetime, leading to the identification of the invariant transverse impact variable with the holographic coordinate $z$. Embedding the light-front Hamiltonian equations in AdS gives rise to semiclassical relativistic bound-state wave equations for arbitrary integer or half-integer spin.  Following the mechanism introduced in~\cite{deAlfaro:1976vlx}, and later extended in Refs.~\cite{Fubini:1984hf} and \cite{Akulov:1983hjq}, the holographic model embodies an underlying superconformal algebraic structure responsible for the introduction of a mass scale $\sqrt{\lambda}$ within the superconformal group, and determines the effective confinement potential for mesons, nucleons and tetraquarks:  It is an effective supersymmetry, not SUSY QCD.  According to Ref.~\cite{deAlfaro:1976vlx} the length scale $1 /\sqrt{\lambda} $ enters the theory through the specific properties of  the vacuum,  the normalizable zero-mass ground state, not through the Lagrangian~\cite{deAlfaro:1976vlx}.  This  model thus  provides important insights into the confinement mechanism of QCD, and, in particular, on the possible origin of the mass scale in QCD from the properties of the nonperturbative vacuum.

There are other aspects and applications of HLFQCD which are not described here but are reviewed in~\cite{Gross:2022hyw}. For example,  LF holographic QCD also incorporates important elements for the study  hadron form factors, such as the connection between the twist of the hadron to the fall-off of its current matrix elements for large $Q^2$, and important aspects of vector meson dominance which are relevant at lower energies. HLFQCD also incorporates features of pre QCD, such as the Veneziano model and Regge theory. Further extensions incorporate the exclusive-inclusive connection in QCD and provide nontrivial relations between hadron form factors and quark distributions~\cite{deTeramond:2018ecg, Liu:2019vsn}, including the intrinsic strange-antistrange~\cite{Sufian:2018cpj} and charm-anticharm~\cite{Sufian:2020coz} asymmetry distributions in the proton. Holographic QCD has also been applied successfully to the description of the gravitational form factors, the hadronic matrix elements of the energy momentum tensor, which provide key information on the dynamics of quarks and gluons within hadrons~\cite{deTeramond:2021lxc}. Holographic QCD has also given new insights on the infrared behavior of the strong coupling in holographic QCD~\cite{Deur:2022msf} and color transparency phenomena~\cite{Brodsky:2022bum}.  Study of diffraction physics~\cite{Forshaw:2012im}  and, most recently, of the EMC effect in various nuclei~\cite{Kim:2022lng}, constitute other interesting examples of the application of the holographic light front ideas to QCD.

\paragraph{Acknowledgments:} I want to thank the organizers of the 25th Bled Workshop, ``What Comes Beyond the Standard Models" for their kind invitation. I am grateful to Stan Brodsky and Hans Guenter Dosch for their invaluable collaboration and to Alexandre Deur, Tianbo Liu, Raza Sabbir Sufian, who have greatly contributed to the new applications of the holographic ideas. 

\end{document}